\documentclass[conference]{IEEEtran}
\IEEEoverridecommandlockouts
\usepackage{cite}
\usepackage{amsfonts}
\usepackage{algorithmic}
\usepackage{graphicx}
\usepackage{textcomp}
\usepackage{xcolor}
\usepackage{multirow}
\usepackage{authblk}
\usepackage{amsmath}
\usepackage{amssymb}
\usepackage{float}
\usepackage{url}
\def\BibTeX{{\rm B\kern-.05em{\sc i\kern-.025em b}\kern-.08em
    T\kern-.1667em\lower.7ex\hbox{E}\kern-.125emX}}

\begin{document}

\title{Deep 3D Vessel Segmentation based on Cross Transformer Network
\thanks{Chengwei Pan and Baolian Qi contribute equally to this work. Corresponding author: Jinpeng Li (lijinpeng@ucas.ac.cn). This work was supported in part by National Natural Science Foundation of China (62106248) and Zhejiang Provincial Natural Science Foundation of China (LQ20F030013).}
}

\author[1]{\textbf{Chengwei Pan}}
\author[2,3]{\textbf{Baolian Qi}}
\author[4]{\textbf{Gangming Zhao}}
\author[1]{\textbf{Jiaheng Liu}}
\author[5]{\textbf{Chaowei Fang}}
\author[6]{\\ \textbf{Dingwen Zhang}}
\author[2,3,*]{\textbf{Jinpeng Li}}
\affil[1]{Institute of Artificial Intelligence, Beihang University, Beijing, China}
\affil[2]{HwaMei Hospital, University of Chinese Academy of Sciences (UCAS), Ningbo, China}
\affil[3]{Ningbo Institute of Life and Health Industry, UCAS, Ningbo, China}
\affil[4]{Department of Computer Science, University of Hong Kong, Hong Kong, China}
\affil[5]{School of Artificial Intelligence, Xidian University, Xi'an, China}
\affil[6]{School of Automation, Northwestern Polytechnical University, Xi'an, China}
\affil[*]{Email: lijinpeng@ucas.ac.cn}

\maketitle

\begin{abstract}
The coronary microvascular disease poses a great threat to human health. Computer-aided analysis/diagnosis systems help physicians intervene in the disease at early stages, where 3D vessel segmentation is a fundamental step. However, there is a lack of carefully annotated dataset to support algorithm development and evaluation. On the other hand, the commonly-used U-Net structures often yield disconnected and inaccurate segmentation results, especially for small vessel structures. In this paper, motivated by the data scarcity, we first construct two large-scale vessel segmentation datasets consisting of 100 and 500 computed tomography (CT) volumes with pixel-level annotations by experienced radiologists. To enhance the U-Net, we further propose the cross transformer network (CTN) for fine-grained vessel segmentation. In CTN, a transformer module is constructed in parallel to a U-Net to learn long-distance dependencies between different anatomical regions; and these dependencies are communicated to the U-Net at multiple stages to endow it with global awareness. Experimental results on the two in-house datasets indicate that this hybrid model alleviates unexpected disconnections by considering topological information across regions. Our codes, together with the trained models are made publicly available at~\url{https://github.com/qibaolian/ctn}.
\end{abstract}

\section{Introduction}

The coronary microvascular disease is a major threat to human health. In the clinic, the key to reducing its impairment is early intervention. Typically, radiologists seek for diagnostic clues for this disease using computed tomography (CT) scans. In recent years, deep learning  algorithms have been developed and applied to help radiologists analyze and diagnose the disease in a more efficient manner, where automatic 3D vessel segmentation is a fundamental step.

Although some excellent works have emerged to tackle this task, some problems remain unsolved. First, there is a lack of carefully annotated datasets to facilitate the proper development and evaluation of algorithms. To our knowledge, the largest open dataset for coronary vessel segmentation was provided by a MICCAI 2020 challenge\footnote{https://asoca.grand-challenge.org/}, which only contained 60 CT volumes. Second, the most popular structure in medical image segmentation, i.e., the U-Net\cite{ronneberger2015u} family, although being data-efficient, needs to be enhanced on the fine-grained vessel segmentation task. Existing U-Nets often yield disconnected and inaccurate predictions, especially for small vessel structures. One reason is that the stacked convolutions are hard to capture the global information and yield topology-ignorant results. Most recently, the transformer-based models have been expected to learn long-range dependencies between anatomical regions. However, compared with the U-Net family, they require larger datasets to train due to the lack of inductive bias, i.e., structural constraints.

A feasible 3D vessel segmentation model should be lightweight, not too demanding on data, and being able to learn 3D global information. This motivates us to explore possibles ways to overlay the advantages of U-Net (\emph{data-efficiency}) and transformer (\emph{global-awareness}). In this design, an important topic is how to interconnect the features of the two types of models to achieve concise and elegant information interaction. 

To implement such design, we propose the cross transformer network (CTN) for fine-grained vessel segmentation. CTN is a hybrid model consisting of a standard U-Net to model the local contextual information and a parallel stacked transformer to learn long-range dependencies between 3D anatomical regions. The dependencies are communicated to the U-Net encoder at multiple stages to endow it with global awareness. Based on this, the model considers both local and global information when segmenting the vessels.

In summary, this paper contributes in two folds:
\begin{itemize}
\item We propose the \textbf{C}ross \textbf{T}ransformer \textbf{N}etwork, a novel method to learn 3D vessel features with global/topology awareness. CTN involves a multi-scale feature interaction between the U-Net and transformer modules, endowing the U-Net with global awareness to confront with disconnections and inaccurate segmentation.
\item We evaluate CTN and existing methods on our two large-scale vessel segmentation datasets. Extensive experiments indicate that CTN achieves state-of-the-art 3D vessel segmentation performance. Our codes and pre-trained models are made publicly available to the community.
\end{itemize}

\section{Related Work}

\subsection{Coronary Artery Disease}
In the last three years, deep learning algorithms have been intensively applied to the automatic diagnosis of coronary artery disease. For example, Denzinger et al. \cite{FelixDenzinger2019CoronaryAP} utilized radiomics, deep learning and their combination to assess the coronary artery plaque segments and the results on 345 plague segments indicated that the combination of shape-, intensity- and texture-based radiomic features with 2D CNN yielded the best result. Denzinger et al. \cite{denzinger2020automatic} further extended the dataset and achieved an accuracy of 0.91 for determining whether the patient suffers from coronary artery disease using a 2D CNN. They used 2D CNN instead of 3D CNN to avoid overfitting. He et al. \cite{JiafaHe2020LearningHR} proposed a hybrid learning algorithm to extract both local and global information, which can guide the automatic extraction of vessel centerlines and confront the discontinuities caused by the segmentation based on local features. Ma et al. \cite{YuxinMa2021SelfSupervisedVS} sequentially used 3D CNNs  and transform modules to capture local and  global information respectively. Then a simple classifier is introduced to predict significant stenosis. In all these works, the centerline extraction is a vital procedure. A well-segmented coronary artery is essential to obtain a reliable centerline and can be further used to analyze vessel morphology to determine angiomas, stenosis, occlusions, etc.

\subsection{CNN-based Segmentation Networks}
Since the introduction of U-Net\cite{ronneberger2015u}, fully convolutional neural networks (CNNs) have been the predominate approach on various 2D and 3D medical image segmentation tasks\cite{dou20163d}\cite{yu2017volumetric}\cite{zhang2020block}\cite{zhu2017deeply}. A vast number of works are dedicated to retrofit and improve the architecture of U-Net, resulting in many variants such as V-Net\cite{milletari2016v}, 3D U-Net\cite{cciccek20163dunet}, Res-UNet\cite{xiao2018weighted}, Dense-UNet\cite{li2018h}, Y-Net\cite{mehta2018net} and  U-Net++\cite{zhou2018unet++}. These methods achieve impressive performance on many tasks, demonstrating the effectiveness of CNN in learning the discriminate features to segment organs or lesions from medical images. Despite their effectiveness, CNNs still suffer from the limited receptive field and lack the ability to capture the long-range (global) dependencies, due to the inductive bias of locality and weight sharing\cite{cohen2016inductive}. Many efforts have been devoted to enlarging the receptive field of CNN with image pyramids\cite{zhao2017pyramid}, atrous convolutions\cite{chen2014semantic} and attention mechanisms\cite{wang2018non}.

\subsection{Vision Transformers}
Transformer architectures using the self-attention mechanism are the most popular methods in natural language processing (NLP) due to their excellent ability of modeling long-range dependencies\cite{vaswani2017attention}. Inspired by this, vision transformers have recently gained traction and achieved competitive performance to CNNs on many computer vision tasks, such as image recognition\cite{kolesnikov2021image}, object detection\cite{dai2021dynamic}, video recognition\cite{girdhar2019video} and semantic segmentation\cite{zheng2021rethinking}. For medical image segmentation, Chen et al.\cite{chen2021transunet} designed the TransUNet for multi-organ segmentation by embedding a transformer as an additional layer in the bottleneck of a U-Net network. Gao et al.\cite{gao2021utnet} proposed a hybrid transformer architecture named UTNet, in which self-attention modules are integrated into both encoders and decoders for capturing long-range dependency at multiple scales. Xie et al.\cite{xie2021cotr} introduced a hybrid model that bridges a CNN and a deformable transformer for 3D medical image segmentation. Similarly, Wang et al.\cite{wang2021transbts} proposed a 3d version of TransUNet, in which a transformer is employed  in the bottleneck of a 3D U-Net architecture for the task of brain tumor segmentation. It is noteworthy that the vision transformers are based on the attention computation and are not specifically designed for the structure of the input data; therefore, a large amount of data is generally required for vision transformers to learn the inductive bias.
\begin{figure*}[t]
\begin{center}
\includegraphics[width=0.9\textwidth]{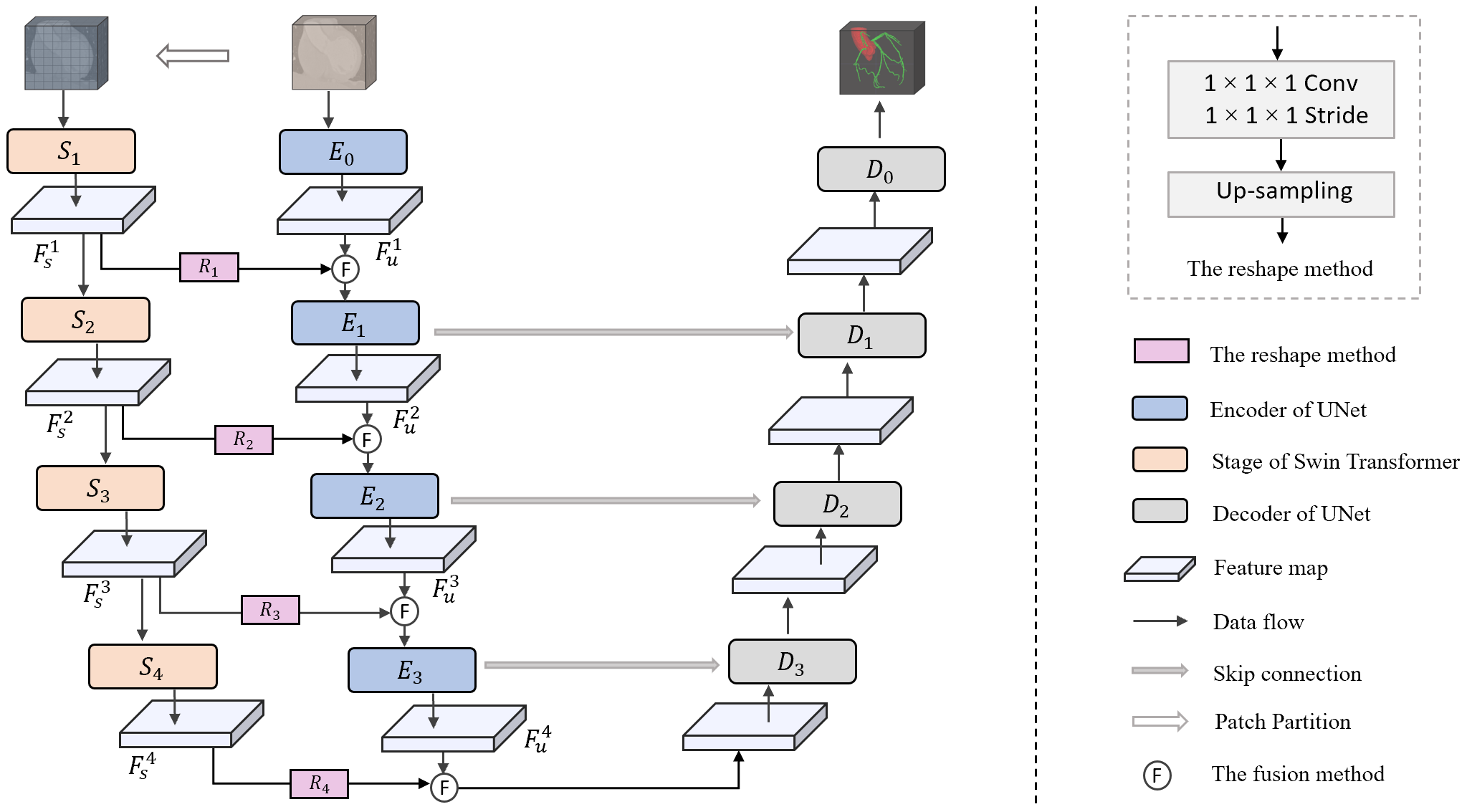}
\caption{Flowchart of the proposed CTN. CT volumes are fed into the 3D U-Net and 3D Swin Transformer to learn local and global features simultaneously. To achieve effective feature interactions, there are four fusions of the U-Net feature map $F_u$ and the Swin Transformer feature map $F_s$, resulting in coase-to-fine feature fusions between the two parallel companions.} \label{fig2}
\end{center}
\end{figure*}
\section{Method}

\subsection{Method}
The architecture of the proposed method is presented in Figure \ref{fig2}. The CT image volume $X\in \mathbb{R}^{D \times H \times W}$ is fed into two branches at the same time, where $D$, $H$, and $W$ represent the spatial depth, height, and width, respectively. The 3D U-Net is the backbone, which includes an encoder and a decoder network. The 3D Swin Transformer acts as a feature extractor to learn the long-term dependencies. In our design, there are four multi-scale fusions of U-Net feature map $F_u$ and Swin Transformer feature map $F_s$. The first two fusions are designed to integrate the coarse-grained representations of the two branches, and the remaining fusions are designed to integrate the fine-grained representations. Note that the size of the reshaped $F_s$ is the same as $F_u$. The aim of this framework is to utilize the long-range contextual information to improve the performance of the vessel segmentation. 



\subsection{3D Shifted Windows}
For the input images $X \in \mathbb{R}^{256\times 256\times 256}$, the windows are partitioned into non-overlapped 3D patches $X'$ with the size of ${4\times 4\times 4}$, and the 3D tokens of $\frac{D}{4} \times \frac{W}{4} \times \frac{H}{4}$ are thus obtained. Due to the large size of CT images, the tiny version of 3D Swin Transformer is adopted in our method. The channel numbers of the hidden layers of four stages are \{48, 96, 192, 384\} and the layer numbers of four stages are \{2, 2, 6, 2\}. Compared to the 2D models, the 3D shifted window based multi-head self-attention is exploited in place of the multi-head self-attention (MSA) module followed by a feed-forward network (FFN) and the other components keep unchanged to those in the 3D Swin Transformer~\cite{liu2021video}.

To reduce computation, the 3D window based MSA (3DW-MSA) module only conducts self-attention within local windows and the computation is implemented as follows.
\begin{equation}
\begin{aligned}
&\hat z^l = 3DW\mbox{-}MSA(LN(z^{l-1}))+z^{l-1},\\
&z^l = FFN(LN(\hat z^l))+\hat z^l,   
\end{aligned}
\end{equation}
where $\hat z^l$ and $z^l$ are the outputs of the 3DW-MSA module and the FFN module in the $l^{th}$ layer. LN denotes layer normalization.

The 3DW-MSA module lacks the information interaction between adjacent windows. Thus, the 3D shifted window based MSA (3DSW-MSA) module is introduced followed by the 3DW-MSA to enlarge the range of dependencies modeled, which is beneficial to the learning of global features. The details of computation are defined as follows.

\begin{equation}
\begin{aligned}
&\hat z^{l+1} = 3DSW\mbox{-}MSA(LN(z^l))+z^{l-1}, \\
&z^{l+1} = FFN(LN(\hat z^{l+1}))+\hat z^{l+1}, 
\end{aligned}
\end{equation}
where $\hat z^{l+1}$ and $z^{l+1}$ are the outputs of the 3DSW-MSA module and the FFN module in the $(l+1)^{th}$ layer. With the shifted windows, the 3D Swin Transformer can better extract feature associations with low computational costs.

\subsection{Fusions in Cross Transformer}
The 3D Swin Transformer is used as an auxiliary encoder of the U-Net to extract long-range dependencies and global contextual connections. After obtaining the output features of the two encoders, it is very important to fuse them effectively and efficiently. There are two fusion methods. One is the adding operation of the feature maps $F_s$ and $F_u$, i.e.,
\begin{equation}
\begin{aligned}
&F^{ij}_{f_{+}} = F^i_u+Reshape(F_s^j), 
\end{aligned}
\end{equation}
where $F^i_u$ is the feature map of the U-Net in the $i^{th}$ encoder and $F^j_s$ is the feature map of the Swin Transformer in the $j^{th}$ stage. $F^{ij}_{f_{+}}$ is the feature map after fusion. The size of feature map $F^j_s$ is reshaped to the same as the feature map $F^i_u$ using the resizing strategies and the $1\times 1$ convolution. 

The other method is concatenating the feature maps $F_s$ and $F_u$ and then performing the convolution operation, i.e.,

\begin{equation}
\begin{aligned}
&F^{ij} = Concat(F^i_u,Reshape(F_s^j) \\
&F^{ij}_{f_{cat}} = Conv(F^{ij}).\\
\end{aligned}
\end{equation}

\section{Experiments}
\begin{figure*}
\begin{center}
\includegraphics[width=0.95\textwidth]{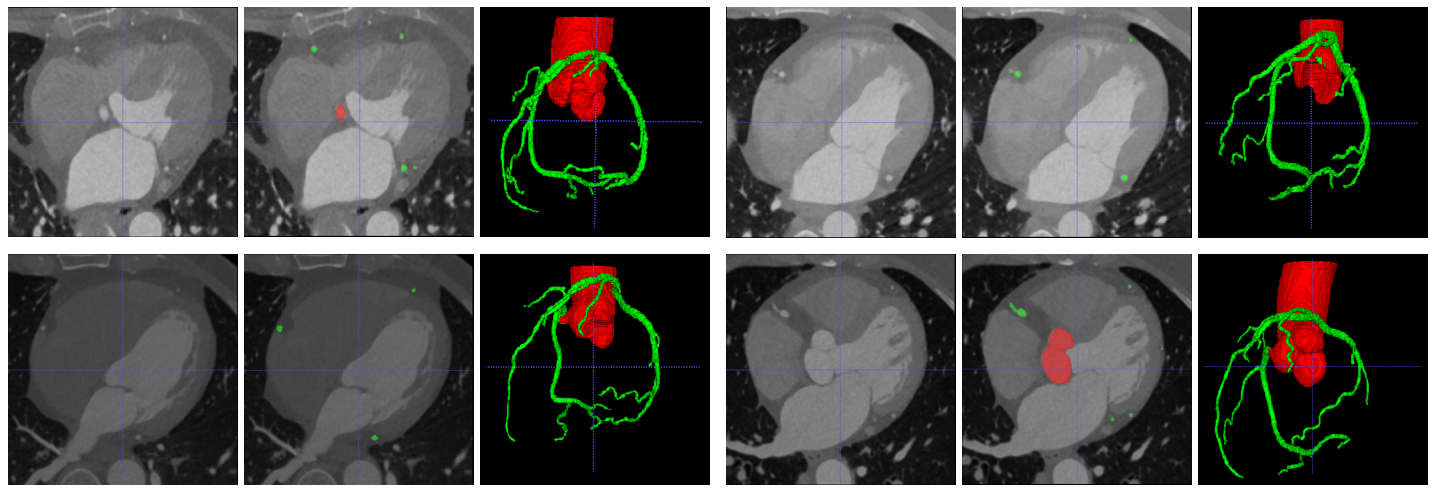}
\caption{Visualization of the ASACA dataset. In each instance, the first column shows the original CT axial images; the second and third columns show images covered with vessel labels and the 3D vessels, respectively. The red and green area mean the aorta and coronary artery, respectively. The aorta is relatively easy to segment, whereas the coronary artery is difficult to segment due to its high diversity and complexity in shape and appearance.} \label{fig01}
\end{center}
\end{figure*}
\subsection{Dataset}
Automated Segmentation of Aorta and Coronary Artery (ASACA) is a large in-house dataset for evaluating the performance of vessel segmentation. Figure \ref{fig01} shows some examples. Each CT image is annotated by one radiologist and verified by another. All the data described in this paper is derived from studies that have received appropriate approvals from institutional ethics committees. The ASACA contains two datasets named ASACA100 and ASACA500, where the main difference is the number of CT images. The ASACA100 consists of a total of 100 coronary computed tomography angiography (CCTA) images, including a training set of 80 CCTA images and a test set of 20 CCTA images. The ASACA500 consists of a total of 500 CCTA images, including a training set of 400 CCTA images, a validation set of 50 CCTA images, and a test set of 50 CCTA images. Please note that all CT images in ASACA are resized to $256\times 256\times 256$. 


\subsection{Experimental Setting}
The model is trained on ASACA500 and ASACA100 using the AdamW optimizer. Within a total of 200 epochs, the learning rate starts from 0.0001 and decreases by 10 times after every 50 epochs. The weight decay and the momentum are set to 0.0001 and 0.9, respectively. The mini-batch size is set to 1 because of the large volume size of the whole 3D CT and the limited GPU memory of a single NVIDIA A100 GPU. The proposed method and all the comparing methods are implemented with PyTorch~\cite{paszke2017automatic}. 
\subsection{Evaluation Metrics}
The dice coefficient (DICE) and average symmetric surface distance (ASSD) are used to evaluate the performance of models, which is commonly used in medical image segmentation~\cite{yue2019cardiac}. Note that the DICE$_A$ denotes the Dice coefficient of aorta and the DICE$_C$ denotes the dice coefficient of coronary artery.
\begin{equation}
DICE= \frac{2|S \cap G|}{|S|+|G|}, 
\end{equation}
where $S$ denotes the segmentation result and $G$ denotes the ground truth. ASSD measures the average symmetrical distance between the segmentation result and the ground truth:
\begin{equation}
\begin{aligned}
&d(G,S)=\mathop{min}_{s \in T(S)}||g-s||, d(S,G)=\mathop{min}_{g \in T(G)}||s-g||,\\
&N=|T(S)|+|T(G)|,\\
&ASSD= \frac{1}{N}(\sum_{g \in T(G)}d(G,S)+\sum_{s\in T(S)}d(S,G)), \\
\end{aligned}
\end{equation}
where $T(S)$ and $T(G)$ denote the set of surface voxels of $S$ and $G$, respectively. $s$ and $g$ are a surface voxel of $T(S)$ and $T(G)$, respectively. $S$ denotes the segmentation result and $G$ denotes the ground truth.

Additionally, the metric of skeleton recall (SR) and skeleton precision (SP) are used to evaluate the tubular structure of coronary artery, which preserve more accurate connectivity of coronary arteries. 
\begin{equation}
\begin{aligned}
SR(S,G)= \frac{|S \cap Q(G)|}{|Q(G)|},   
SP(S,G)= \frac{|G \cap Q(S)|}{|Q(S)|}, 
\end{aligned}
\end{equation}
where $S$ and $G$ are the segmentation result and the ground truth, respectively. Q(*) is the skeletonization function~\cite{van2014scikit}, which has been used to preserve the original vascular topology and connectivity.
\begin{table*}
\begin{center}
\caption{The performance comparison of vessel segmentation among the models using ASACA500 and ASACA100 datasets. Note that mm denotes millimeters and $_{(128)}$ denotes the input resolution of ($128 \times 128 \times128$).} \label{tab1}

\resizebox{0.92\textwidth}{!}{
\begin{tabular}{ l|c| c| c| c| c| c| c}
\hline
 Dataset & Models & DICE(\%)& DICE$_A$(\%)& DICE$_C$(\%)& ASSD(mm) & SP(\%) & SR(\%) \\
\hline
\multirow{7}{*}{ASACA500}& UNet~\cite{cciccek20163dunet} & 91.57 & 98.03 & 85.15 & 0.479 & 95.43 & 91.82  \\						
								& Swin Transformer~\cite{liu2021video} & 89.61 & 97.41 & 82.14 & 0.633 & 94.44 & 87.88  \\
								& clDice~\cite{shit2021cldice} & 91.71 & 98.04 & 85.46 & 0.466 & 95.60 & 92.40  \\
								& TransUNet~\cite{chen2021transunet} & 90.89 & 97.86 & 83.96 & 0.546 & 96.49 & 88.39  \\
								& UTNet$_{(128)}$~\cite{gao2021utnet} & 86.50 & 98.01 & 85.37 & 0.497 & 94.67 & 92.25  \\
								& CoTr$_{(128)}$~\cite{xie2021cotr} & 84.87 & 97.68 & 83.72 & 0.605 & 95.73 & 88.43  \\
								& Ours & \textbf{91.89} & \textbf{98.07} & \textbf{85.79} & \textbf{0.429} & \textbf{95.75} & \textbf{93.20}  \\

\hline
\multirow{7}{*}{ASACA100}& UNet~\cite{cciccek20163dunet} & 89.93 & 97.13 & 83.21 & 0.616 & 94.00 & 90.35  \\						
								& Swin Transformer~\cite{liu2021video} & 86.43 & 94.72 & 78.15 & 1.299 & 94.23 & 78.38  \\
								& clDice~\cite{shit2021cldice} & 90.27 & 97.26 & 83.36 & 0.656 & \textbf{94.97} & 89.30  \\
								& TransUNet~\cite{chen2021transunet} & 88.48 & 96.50 & 80.75 & 0.994 & 93.93 & 84.28  \\
								& UTNet$_{(128)}$~\cite{gao2021utnet} & 83.48 & 97.17 & 84.38 & 0.663 & 94.00 & 91.24  \\
								& CoTr$_{(128)}$~\cite{xie2021cotr} & 77.85 & 96.32 & 80.03 & 1.016 & 91.10 & 83.89 \\
								& Ours & \textbf{91.01} & \textbf{97.33} & \textbf{84.93} & \textbf{0.499} & 94.77 & \textbf{93.14} \\
\hline

\end{tabular}}
\end{center}
\end{table*}

\begin{table*}[h]
\begin{center}
\caption{The performance comparison of different fusion methods using ASACA500 and ASACA100 dataset. Note that mm denotes millimeters. }\label{tab2}

\resizebox{0.92\textwidth}{!}{
\begin{tabular}{ l|c| c| c| c| c| c| c}
\hline
 Dataset & Models & DICE(\%) & DICE$_A$(\%) & DICE$_C$(\%) & ASSD(mm) & SP(\%) & SR(\%) \\
\hline
\multirow{2}{*}{ASACA500}& $F_{f_{cat}}$ & 91.87 & 98.04 & \textbf{85.85} & 0.433 & \textbf{95.77} & \textbf{93.35}  \\						
								& $F_{f_+}$ & \textbf{91.89}  & \textbf{98.07} & 85.79 & \textbf{0.429} & 95.75 & 93.20  \\

\hline
\multirow{2}{*}{ASACA100}& $F_{f_{cat}}$ & 90.54 & 97.29 & 84.08 & 0.597 & 94.32 & 92.07  \\						
								& $F_{f_+}$ & \textbf{91.01} & \textbf{97.33} & \textbf{84.93} & \textbf{0.499} & \textbf{94.77} & \textbf{93.14} \\
\hline
\end{tabular}}
\end{center}
\end{table*}

\begin{table*}[h]
\begin{center}
\caption{The performance comparison of different fusion stages using ASACA500 and ASACA100 dataset. Note that mm denotes millimeters. }\label{tab3}

\resizebox{0.92\textwidth}{!}{
\begin{tabular}{ l|c c c c| c| c|c| c| c| c}
\hline
 Dataset & $F^{11}_f$ & $F^{22}_f$ & $F^{33}_f$ & $F^{44}_f$ & DICE(\%) & DICE$_A$(\%) & DICE$_C$(\%)& ASSD(mm) & SP(\%) & SR(\%) \\
\hline
\multirow{6}{*}{ASACA500}& $\boxtimes$ & $\boxtimes$ & $\boxtimes$ & $\boxtimes$ & 91.57  & 98.03 & 85.15 & 0.479 & 95.43 & 91.82  \\						
								& $\checkmark$ & $\checkmark$ & $\checkmark$ & $\boxtimes$ & 91.79 & 98.04 & 85.63 & 0.430 & 95.73 & 93.44  \\
								& $\checkmark$ & $\checkmark$ & $\boxtimes$ & $\checkmark$ & 91.80 & 98.04 & 85.76 & 0.426 & 94.83 & 94.60  \\
								& $\checkmark$ & $\boxtimes$ & $\checkmark$ & $\checkmark$ & 91.77 & 98.04 & 85.59 & 0.444 & 96.35 & 92.15  \\
								& $\boxtimes$ & $\checkmark$ & $\checkmark$ & $\checkmark$ & 91.83 & 98.07 & 85.65 & 0.440 & 96.59 & 92.42   \\
								& $\checkmark$ & $\checkmark$ & $\checkmark$ & $\checkmark$ & 91.89 & 98.07 & 85.79 & 0.429 & 95.75 & 93.20  \\

\hline
\multirow{6}{*}{ASACA100} & $\boxtimes$ & $\boxtimes$ & $\boxtimes$ & $\boxtimes$ & 89.93 & 97.13 & 83.21 & 0.616 & 94.00 & 90.35  \\						
								& $\checkmark$ & $\checkmark$ & $\checkmark$ & $\boxtimes$  & 90.83 & 97.30 & 84.74 & 0.535 & 93.29 & 94.22  \\
								& $\checkmark$ & $\checkmark$& $\boxtimes$ & $\checkmark$  & 90.85 & 97.20 & 84.62 &  0.530 & 94.72 & 92.11  \\
								& $\checkmark$ & $\boxtimes$ & $\checkmark$ & $\checkmark$ & 90.87 & 97.18 & 84.62 & 0.538 & 95.69 & 91.33  \\
								& $\boxtimes$ & $\checkmark$ & $\checkmark$ & $\checkmark$ & 90.87 & 97.31 & 84.56 & 0.535 & 94.27 & 92.81  \\
								& $\checkmark$ & $\checkmark$ & $\checkmark$ & $\checkmark$ & 91.01 & 97.33 & 84.93 & 0.499 & 94.77 & 93.14 \\
\hline
\end{tabular}}
\end{center}
\end{table*}

\subsection{Comparison with the Reference Model}

To evaluate the effectiveness of the proposed model in a fair manner, we reproduce and compare six representative methods on ASACA500 dataset and ASACA100 dataset under the same data partitioning and experimental setup. The six compared methods consist of the 3D U-Net~\cite{cciccek20163dunet}, 3D Swin Transformer~\cite{liu2021video}, clDice~\cite{shit2021cldice}, TransUNet~\cite{chen2021transunet}, UTNet~\cite{gao2021utnet} and CoTr~\cite{xie2021cotr}. The 3D U-Net~\cite{cciccek20163dunet} is the 3D version of U-Net~\cite{OlafRonneberger2015UNetCN}, where the 2D operations of which are all replaced 
by their 3D counterparts for volumetric segmentation. 3D Swin Transformer~\cite{liu2021video} is the 3D version of Swin Transformer~\cite{ZeLiu2021SwinTH}, which extends the scope of local attention computation from only the spatial domain to the spatiotemporal domain. The clDice~\cite{shit2021cldice} is the state-of-the-art method for tubular structure segmentation, which guarantees topology preservation up to homotopy equivalence for binary 2D and 3D segmentation. The TransUNet\cite{chen2021transunet} embeds a transformer as an additional layer in the bottleneck of a U-Net network for multi-organ segmentation. UTNet \cite{gao2021utnet} is a hybrid transformer architecture, in which self-attention modules are integrated in both encoders and decoders for capturing long-range dependency at multiple scales. CoTr \cite{xie2021cotr} is a hybrid model that bridges a CNN and a deformable transformer for 3D medical image segmentation. Especially, UTNet and CoTr have a low input resolution ($128 \times 128 \times128$) because of the GPU memory constraint, and the other methods have a default input resolution of $256\times 256 \times256$ as described in the Section 3.1. 


The experimental results in Table \ref{tab1} show that our model achieves state-of-the-art performance in six evaluation metrics on the two datasets. For example, our model achieves the DICE of 91.89\%, the ASSD of 0.429mm, the SP of 95.75\% and the SR of 93.20\% on ASACA500 dataset, and the DICE outperforms U-Net, Swin transformer, clDice, TransUNet, UTNet, and CoTr by 0.32\%, 2.28\%, 0.18\%, 1.00\%, 5.39\%, and 7.02\%, respectively. It is noteworthy that UTNet and CoTr achieve poor performance because of low input resolution compared with the other methods. Additionally, since aorta is relatively easy to segment and contributes high performance, the Dice of aorta (DICE$_A$) and the DICE of coronary artery (DICE$_C$) are reported separately. Overall, the experimental results demonstrate that the proposed CTN can efficiently represent sparse and anisotropic vessel structures and has a good performance for vessel segmentation.
\begin{figure*}
\begin{center}
\includegraphics[width=1.0\textwidth]{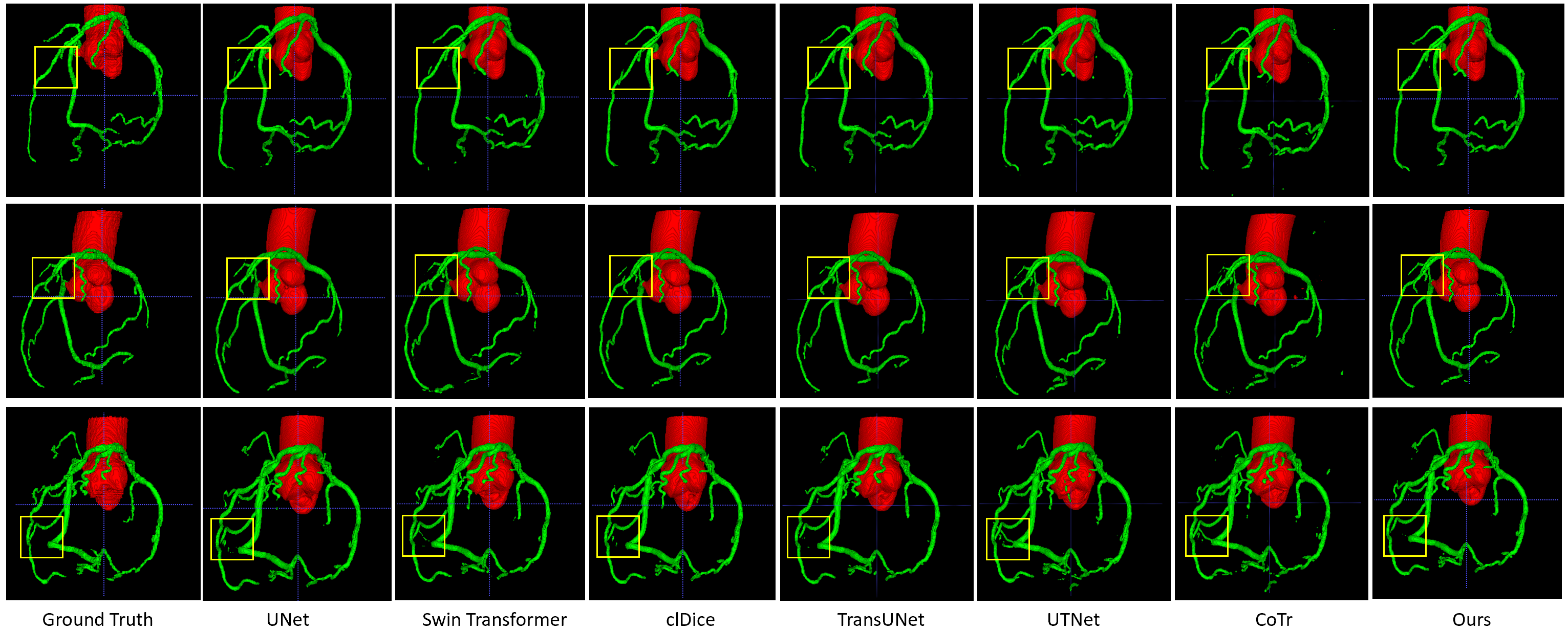}
\caption{Visualization of the vessel segmentation of different models, including the ground truth, the results of U-Net, Swin Transformer, clDice, TransUNet, UTNet, CoTr, and our model (from left to right). The red and green areas mean the aorta and the coronary vessels. The proposed CTN makes more accurate and continuous predictions, which is in consistent with the quantitative results.} \label{fig03}
\end{center}
\end{figure*}

\section{Ablation Study}

\subsection{Ablation of Fusion Method}
We investigate the effectiveness of the two fusion methods, including the addition operation $F_{f_+}$ and the concatenation operation $F_{f_{cat}}$. As is shown in Table \ref{tab2}, the performance of the fusion method $F_{f_+}$ is better than that of the fusion method $F_{f_{cat}}$ on ASACA500 and ASACA100 dataset. Especially, the fusion method $F_{f_{+}}$ has more obvious advantages on ASACA100 dataset. For example, $F_{f_{+}}$ outperforms $F_{f_{cat}}$ by 0.02\% (DICE) on ASACA500 dataset, but $F_{f_{+}}$ outperforms $F_{f_{cat}}$ by 0.47\% (DICE) on ASACA100 dataset. The reason may be that the concatenation operation requires more data to learn a good weight of feature vectors, while the addition operation does not. Therefore, the fusion method $F_{f_{+}}$ is used to fuse features in our experiment. 

\subsection{Ablation of Fusion Stages}
We also investigate the effectiveness of different fusion stages. We discard different fusion stages and evaluate the effect of each fusion stage. According to Table \ref{tab3}, the models with different fusion stages demonstrate more advantages over the baseline model and the model with all fusion stages achieves the best performance. It can be seen that all fusion stages play important roles to improve segmentation performance. Moreover, removing $F^{33}_f$ leads to 0.15\% (DICE) performance drop and removing $F^{44}_f$ leads to 0.18\% (DICE) performance drop on ASACA100 dataset. Additionally, if the fusion of $F^{33}_f$ and $F^{44}_f$ is discarded, the model suffers from more performance drop. Therefore, $F^{22}_f$, $F^{33}_f$, and $F^{44}_f$ are the most important communication in the model. It demonstrates that the extraction of global information is necessary for vessel segmentation, which is consistent with our motivation to take advantage of the ability of transformer modules to capture long-range dependencies. These observations are inconsistent with our motivation.

\subsection{Visualization}
We visualize some typical predictions of our model and the compared methods in Figure \ref{fig03}. The proposed CTN predicts more accurate segmentation results, and other models achieve relatively low performance due to over-segmentation or under-segmentation in high-density regions. It demonstrates that the method of the multi-scale feature interaction between the U-Net and transformer modules can further boost the segmentation performance.

\section{Conclusion}
We exploit and explore the CTN to better segment coronary arteries by addressing the insufficiency of existing 3D U-Nets, i.e., the neglect of global structural information. In practice, this can partially lead to discontinuous and inaccurate segmentation results. Comparison experiments and ablation experiments on two largest datasets to date demonstrate that our hybrid model is superior to state-of-the-art. The CTN is designed to better take into account global topological information to combat disconnections and inaccurate segmentation results. Qualitative and quantitative results indicate that the global features learned by transformers are able to compensate for the deficiency of the U-Net structure, i.e., the topology-ignorance. The long distances encourage the continuity of tiny vessel structures. Our implementation will serve as a strong baseline in the CT-based vessel segmentation. In future work, we will explore alternative ways of interacting global-local information between parallel models and investigate lightweight model implementations.

{
\bibliographystyle{IEEEtranS}
\bibliography{egbib}
}
\end{document}